\documentclass[a4paper,fleqn,usenatbib]{mnras}

\usepackage{mathptmx}

\usepackage[T1]{fontenc}
\usepackage{ae,aecompl}
\usepackage{etoolbox}
\usepackage[dvipsnames]{xcolor}

\makeatletter
\patchcmd\@combinedblfloats{\box\@outputbox}{\unvbox\@outputbox}{}{%
   \errmessage{\noexpand\@combinedblfloats could not be patched}%
}%
 \makeatother

\usepackage{graphicx}	
\usepackage{amsmath}	
\usepackage{amssymb}	

\newcommand{\ha}{H$\alpha$}
\newcommand{\hb}{H$\beta$}
\newcommand{\hg}{H$\gamma$}
\newcommand{\hd}{H$\delta$}

\newcommand{\hei}{He~{\sc i}}
\newcommand{\heii}{He~{\sc ii}}

\def\kms{\mbox{${\rm km}\:{\rm s}^{-1}\:$}}

\def\lesssim{\mathrel{\hbox{\rlap{\hbox{\lower4pt\hbox{$\sim$}}}\hbox{$<$}}}}
\let\la=\lesssim
\def\gtrsim{\mathrel{\hbox{\rlap{\hbox{\lower4pt\hbox{$\sim$}}}\hbox{$>$}}}}
\let\ga=\gtrsim

\def\sol{$~\mathrm{M}_\odot$}

\def\ergs{erg s$^{-1}$}

\title[The accretion disc wind of V4641~Sgr]{The low-luminosity accretion disc wind of the black hole transient V4641~Sagittarii}

\author[Mu\~noz-Darias, Torres \& Garcia]{Teo Mu\~noz-Darias$^{1,2}$, Manuel~A.~P.~Torres $^{1,2,3}$, Michael~R.~Garcia$^{3}$\\
$^{1}$ Instituto de Astrof\'isica de Canarias, 38205 La Laguna, Tenerife, Spain \\
$^{2}$ Departamento de astrof\'isica, Univ. de La Laguna, E-38206 La Laguna, Tenerife, Spain \\
$^{3}$ SRON Netherlands Institute for Space Research Sorbonnelaan 2, 3584 CA Utrecht, The Netherlands \\
$^{4}$ NASA Headquarters, Mail Suite 3Y28, 300 E Street SW, Washington, DC 20546-0001, USA \\
}
\date{Accepted XXX. Received YYY; in original form ZZZ}

\pubyear{2018}

\begin{document}
\label{firstpage}
\pagerange{\pageref{firstpage}--\pageref{lastpage}}
\maketitle

\begin{abstract}
We present an optical spectroscopic study of the black hole X-ray transient V4641 Sgr (=SAX~J1819.3-2525) covering the 1999, 2002 and 2004 outbursts. The spectra were taken over 22 different epochs during the low luminosity phases that follow the sharp and bright outburst peaks displayed by the system. The data reveal the frequent presence of wind-related features in H (Balmer) and  \hei\ emission lines in the form of P-Cygni profiles and strong emission lines with broad wings. The terminal velocity of the wind, as measured in the blue-shifted absorption (P-Cygni) components, is  in the range of $\sim$ 900--1600 \kms, while the broad emission line wings (so-called nebular phases)  imply outflow velocities of  up to $\sim 3000$ \kms.  We show that, at least for several of the wind detections, the radio jet was active and the system was likely in the hard state. This, together with previous detections reported in the literature, shows that V4641 Sgr is the second source of this class, after V404 Cyg, where the presence of these cold wind outflows has been clearly established. We discuss the similar phenomenology observed in both systems as well as the possible nature of the outflow and its impact on the accretion process.   
\end{abstract}



\defcitealias{Munoz-Darias2016}{MD16} 
\section{Introduction}

Black hole X-ray transients (BHTs) are X-ray binary systems formed by a Roche-lobe filling star transferring matter onto a stellar-mass black hole. BHTs, as a class, spend most part of their lives in dim, quiescent states, showing occasional outbursts when their X-ray luminosity increases by several orders of magnitude, rivalling in some cases with the brightest sources of the night X-ray sky. These activity periods, which typically last several months, are characterized by a well defined phenomenology  of ``X-ray states" \citep{McClintock2006, Belloni2011} accompanied by outflows in the form of radio-jets \citep{Corbel2002, Gallo2003, Fender2004} and hot accretion disc winds detected as blue-shifted X-ray absorptions produced by highly ionised material. The latter are solely seen in high inclination systems, which is taken as a proxy of an equatorial geometry \citep{Ponti2012}. These show terminal velocities in the range of $\sim$ 100--2000~\kms and are typically observed in the so-called \textit{soft states}, when the radio jet is not detected  \citep{Miller2006,Neilsen2009,Ponti2012,Ponti2016}. The \textit{hard state}, on the other hand, is characterised by radio emission from a compact, unresolved jet; while discrete ejections are commonly observed during the transition between the two aforementioned X-ray states. This rather complex accretion-ejection coupling \citep[][for a review]{Fender2016} is mostly shared by the larger sample of low-mass X-ray binaries with neutron star accretors \citep{Munoz-Darias2014}, which also show strong radio-jets and hot winds \citep{Migliari2006,DiazTrigo2006,DiazTrigo2016,Ponti2014,Ponti2015}. 

This picture became even more complicated after the luminous and violent June 2015 outburst of the BHT V404 Cyg. During this event,  conspicuous P-Cygni profiles were discovered in a dozen of recombination optical lines of He and H, revealing the presence of an outflowing accretion disc wind with a terminal velocity in the range of 1500--3000 ~\kms. The wind was strictly simultaneous with the radio jet \citep[][hereafter \citetalias{Munoz-Darias2016}]{Munoz-Darias2016} and contemporaneous with a complex, X-ray wind (\citealt{King2015}; see also \citealt{Motta2017a}). This cold outflow was observed also in the form of an optically thin phase (so-called \textit{nebular phase}; \citetalias{Munoz-Darias2016}) characterized by unprecedentedly strong and very broad \ha\ emission (equivalent width of $\sim 2000$ \AA\ and full width at zero intensity of $\sim$ 6000 \kms). Wind features were also present in optical spectra taken during the 1989 outburst and during the fainter December 2015 sequel outburst \citep{Casares1991,Munoz-Darias2017},  suggesting that this type of outflow is most likely present in every V404 Cyg outburst event. Additional evidence for the simultaneous presence of winds and jets in low mass X-ray binaries was reported by  \citet{Homan2016}.

V4641 Sagittarii (V4641~Sgr hereafter; a.k.a. SAX~J1819.3-2525) is an X-ray binary system formed by a (late) B-type donor transferring mass onto a $\sim 6$ \sol\ black hole. The binary has an orbital period of 2.8 d and it is placed at a distance of  $\sim 6.2$ kpc \citep{Orosz2001, MacDonald2014}. The latter value is consistent with the parallax measurement reported in the \textit{Gaia data release 2} \citep{Gaia2018}. These properties make V4641~Sgr a rather peculiar system within the known population of  BHTs \citep{Casares2014,Corral-Santana2016}. In particular, it has the most massive donor star ($\sim$ 3\sol; \citealt{MacDonald2014}), which largely contributes to the total optical flux during outburst and fully dominates in quiescence with $V\sim13.7$ \citep{Orosz2001}. It has also one of the longest orbital periods and therefore accretion discs after GRS~1915+105 (which is showing a quasi-persistent behaviour) and V404 Cyg itself. As noted in \citetalias{Munoz-Darias2016},  V4641~Sgr is,  within the known population of BHTs, the one which has displayed optical phenomenology more similar to that of V404 Cyg, with broad \ha\ emission and weak P-Cygni profiles in Fe~{\sc ii} at 5169 \AA\ with a terminal velocity of $\sim$ 600 \kms \citep{Lindstrom2005,Chaty2003}. In this paper we present an optical spectroscopic study carried out over the 1999 discovery outburst \citep{Zand2000}, as well as the 2002 and 2004 events. We report on the presence of wind-related features in each of these outburst events, suggesting an analogous behaviour to that observed in V404 Cyg. 
  
\section{Observations}
A total of 67 optical spectra of V4641 Sgr were obtained using the FAST spectrograph \citep{Fabricant1998} attached  to the 1.5m telescope on the Fred L. Whipple Observatory (Mount Hopkins, Arizona, USA). Among them, 39 were taken over 22 different epochs during the 1999, 2002 and 2004 outbursts. In addition, 28 spectra were taken during quiescence and presented in the dynamical study by \citet{Orosz2001}.  A log of the observations is given in Table \ref{log}. The spectroscopy was carried out using  the 300 g mm$^{-1}$ grating with a 2720 x 512 pix$^2$ CCD. Spectra cover the wavelength range of  3500 to 7700 \AA\ and were taken with different long slits; the majority using the 1.1 and 1.5 arcsec slits,  while a few spectra were acquired using  wider widths (3 and 5 arcsec). The seeing was in the range of 1 to 2 arcseconds during the observations. Standard data reduction consisting in  bias subtraction, flat fielding and wavelength calibration was performed to the two-dimensional CCD frames using \textsc{iraf}. One-dimensional spectra were optimally extracted and subsequently normalised and analysed using \textsc{molly} software and custom routines based in {\sc python}. Using the sky line [O\textsc{i}] 5577.338 \AA\ we determined the nominal spectral resolution, which is in the range of 240--320 \kms. Likewise, the wavelength calibration was found to be accurate to, at least, 10 per cent of the aforementioned resolution.
  
\begin{table}
	\centering
	\caption{Observing log. MJD corresponds to the mean time of the first exposure. Data in boldface correspond to observations where wind-related features are present in the spectra.}
	\label{log}
	\begin{tabular}{cccc} 
		\hline
		Date & MJD &  Exposure Time (s)  & State \\
		\hline
		
		\textbf{17/09/1999} & 51438.1871 & $7 \times 600$                                  & Outburst \\   
       18/09/1999 & 51439.1089 & $2 \times 300$ + $10 \times 600$  & Outburst\\	                 
        \hline
       19/09/1999 & 51440.1021 & $90$ + $600$ +$4 \times 300$ & quiescence$^{*}$ \\
		02/10/1999 & 51453.0842 &  $180$ + $4 \times 240$ & quiescence \\
		03/10/1999 & 51454.1064 & $4 \times 240$ & quiescence \\
		04/10/1999 &51455.1168 & $3 \times 240$ & quiescence \\
		11/10/1999 & 51462.0882 & $2 \times 240$ & quiescence \\
		14/10/1999 & 51465.1022 & $2 \times 600$ & quiescence \\
		15/10/1999 & 51466.0874 &  $3 \times 600$ & quiescence \\
		16/10/1999 & 51467.0786 &  $3 \times 720$  & quiescence \\
		\hline		
		05/06/2002 & 52430.3797 & 1200 & Outburst \\						
		10/06/2002 & 52435.4023 & 1200 & Outburst\\						
		11/06/2002 & 52436.3939 & 1200 & Outburst \\						
		\textbf{12/06/2002} & 52437.3425 & 1200 & Outburst \\			
		\textbf{14/06/2002} & 52439.3562 & 1200 &  Outburst \\			
		\textbf{15/06/2002} & 52440.3560 & 1200 &  Outburst \\			
		\textbf{17/06/2002} & 52442.3854 & 1200 &  Outburst \\						
		\textbf{18/06/2002} & 52443.3633 & 1200 &  Outburst \\			
		\textbf{19/06/2002} & 52444.3541 & 1200 &  Outburst \\			
		\textbf{20/06/2002} & 52445.3416 & 1200 &  Outburst \\			
		\textbf{04/07/2002} & 52459.2899 & 1200 &  Outburst\\			
		\textbf{05/07/2002} & 52460.2795 & 1200 &  Outburst\\			
		\textbf{06/07/2002} & 52461.2764 & 1200 &  Outburst\\			
		\textbf{07/07/2002} & 52462.2874 & 1200 &  Outburst \\						
		\textbf{10/07/2002} & 52465.3634 & 1200 &  Outburst\\			
		\textbf{11/07/2002} & 52466.2880 &   960 &  Outburst\\			
		12/07/2002 & 52467.3024 & 1200 & Outburst \\						
		18/07/2002 & 52473.3101 & 1200 & Outburst\\						
		\hline
		\textbf{19/07/2004 }& 53205.2475 & 600 &  Outburst \\			
        \textbf{21/07/2004} & 53207.2390 & 600 & Outburst \\				
		\hline 
	\end{tabular}
\begin{flushleft}
*some enhanced activity is present in the optical spectra (see Section \ref{quiescence})
\end{flushleft}
\end{table}

\begin{figure*}
\begin{center}
\includegraphics[keepaspectratio,width=15.cm]{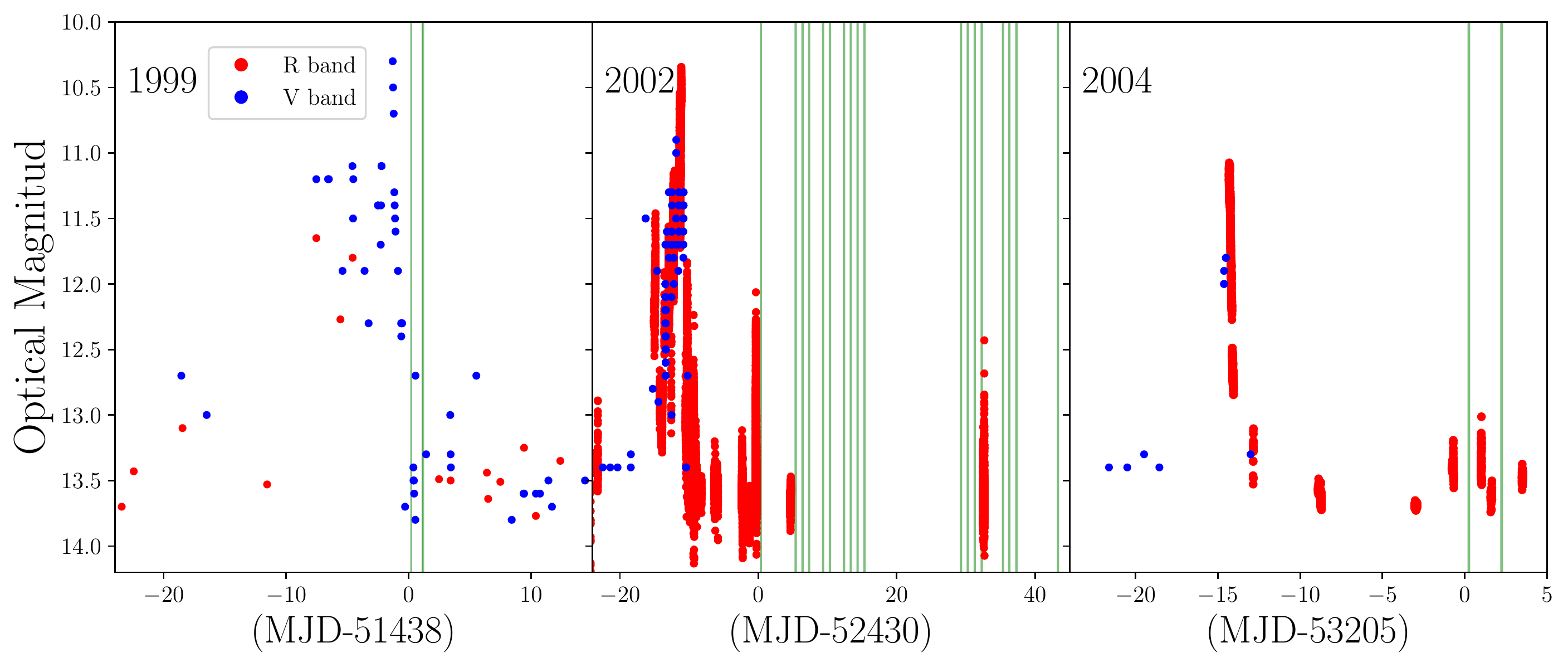}
\caption{Photometric database collected during the 1999, 2002, and 2004 outbursts and presented in \citet{Uemura2002, Uemura2004b, Uemura2005}, respectively. Red dots correspond to R-band data, while blue stars represent V-band observations gathered by VSNET. Green, vertical lines mark the times when optical spectra were taken (see Table \ref{log}). In each panel the zero time is set by the day of the first spectrum. }
\label{fig:lc}
\end{center}
\end{figure*}

\subsection{The quiescence average spectrum}
\label{quiescence}
While the goal of the paper is to search for wind-related signatures in optical lines during outburst episodes, a good characterization of the companion star optical spectrum is particularly useful given that the stellar absorption features are present in the outburst spectra, owing to the strong contribution of the B-type donor to the optical flux. 
As a first step, we collected the 28 normalized spectra taken with the same instrumental set-up across 8 epochs in September and October 1999 (see Table \ref{log}) and combined then into a single, high signal-to-noise average spectrum. To carry out this task we first removed from every spectrum the orbital radial velocity of the donor at the time of the observation using the  orbital solution  by  \citet{Orosz2001}.  

In a second step, we divided each of the 28 quiescent spectra by the average spectrum shifted to the corresponding donor radial velocity at the time of each observation. For the 22 spectra taken in October 1999 (over 7 different epochs; see Table \ref{log}), this yielded a spectrum totally flat with some occasional noisy and unstructured residuals at the positions of the Balmer absorption lines. In contrast, the 6 donor-normalized spectra taken in September 19, 1999  revealed the presence of weak, but clearly detected, residuals in the form of Balmer emission lines (see section \ref{analysis}). This is not surprising given that the system was nominally in outburst one day before (see below and  \citealt{Orosz2001}). 
In light of this finding, we repeated the first step and computed the average donor spectrum solely using the October 1999 data, to which we will refer hereafter as the \textit{donor spectrum}.

\begin{figure*}
\begin{center}
\includegraphics[keepaspectratio,width=15.cm]{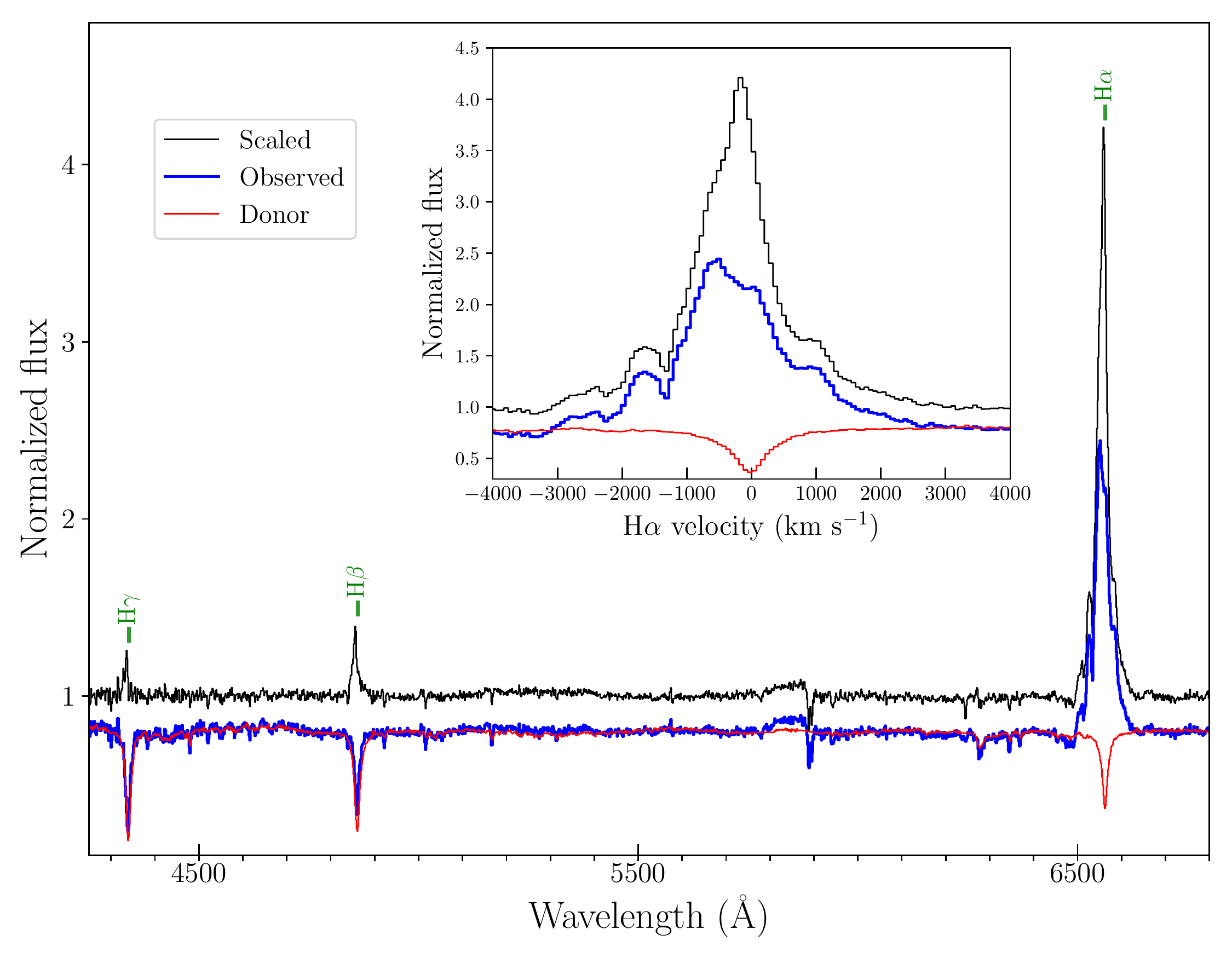}
\caption{Average spectrum obtained during the 1999 outburst (Sep 17; blue line) just $\sim$ 1.5 days after the flux peak and following a 3-mag drop.  The average, quiescent spectrum, which is fully dominated by the donor, is shown in red, while the donor-scaled spectrum (see text) is represented as a black, solid line. Spectra are displayed normalized and in the rest frame of the donor. The outburst (blue) and quiescence (red) spectra are offset by -0.2 in the Y-scale.
The inset shows the \ha\ region, where a complex line profile, including emission line wings extending up to $\sim \pm 3000$ \kms can be appreciated. The absorption troughs observed in the blue wing are most likely not related to the source (see section \ref{1999}). }
\label{fig:1999}
\end{center}
\end{figure*}

\subsection{Outburst data}
\label{odata}
In this work we use 39 optical spectra taken when V4641~Sgr was in outburst. For the 2002 and 2004  data, only one spectrum per epoch was taken, while several consecutive spectra were obtained in each of the two 1999 observing nights (Table \ref{log}). After a careful inspection, we observed clear night-to-night variations but no significant changes are observed within each of the two 1999 observing nights. Thus, spectra were nightly averaged. Hence, in our analysis we finally considered 22 single-epoch spectra: 2 from 1999, 18 from 2002 and 2 taken during the 2004 outburst. 

As noted above, the donor star strongly contributes to the optical emission of the system during outburst. Indeed, the Balmer lines, with the exception of \ha\ are  always detected in absorption, while the emission line profiles characteristic of BHTs in outburst are only occasionally  seen in \hei\ and \heii\ transitions. However, clear variations in the  depth of the Balmer lines are observed from epoch to epoch. In particular, the \ha\ emission dramatically varies, showing in several cases bizarre line profiles resulting from the interplay between the absorptions associated with the companion star lines and the accretion-related emission that fill them in. 

To tackle the above problem, we divided each spectrum by the \textit{donor spectrum}. As we did before for quiescent data (see section \ref{quiescence}), this was carried out by considering the donor star radial velocity at the time that each single spectrum was taken. The resulting \textit{donor-scaled} database requires some important remarks: (i) it is specially useful when dealing with spectral regions dominated by the donor emission or when it represents a significant contribution  (i.e. at least every Balmer line but \ha). However, as we are working with normalized spectra, we must be very cautious when extracting conclusions from these regions, since the result of just having extra emission with no particular shape (e.g. a flat accretion disc contribution) would produce spurious emission lines. (ii) this method assumes that the spectrum of the donor does not change from quiescence to outburst, as e.g. result of strong X-ray irradiation. This assumption is favoured by the low luminosity at which our observations were taken (see below) and by the early spectral type of the donor star. However, we note that minor differences were found between the inner (i.e. irradiated) and outer (i.e. non-irradiated) donor faces of the B-type companion in LMC-X-3 \citep[see][and references therein]{Orosz2014}, albeit this system has a shorter orbital period and is persistently accreting at luminosities $\gtrsim 10^{37}$ \ergs. (iii) The spectral regions where the donor spectrum is featureless are unaffected.  (iv) Special attention needs to be put on spectral regions dominated by emission not related to the donor (e.g. \ha). On the other hand, the resulted donor-scaled database tends to be qualitatively more similar to that typical of BHTs in outburst, with Balmer emission in every transition of the series and more usual \ha\ emission profiles. In the following sections, we will study separately the different observations taken in every outburst by initially considering the results of the regular database (\textit{observed} database hereafter), while the donor-scaled spectra will be used to provide further support to the inferred conclusions. We shall note that all the values reported in the paper (e.g. equivalent widths) are derived from the observed database unless it is stated otherwise.

\begin{figure*}
\begin{center}
\includegraphics[keepaspectratio,width=15.cm]{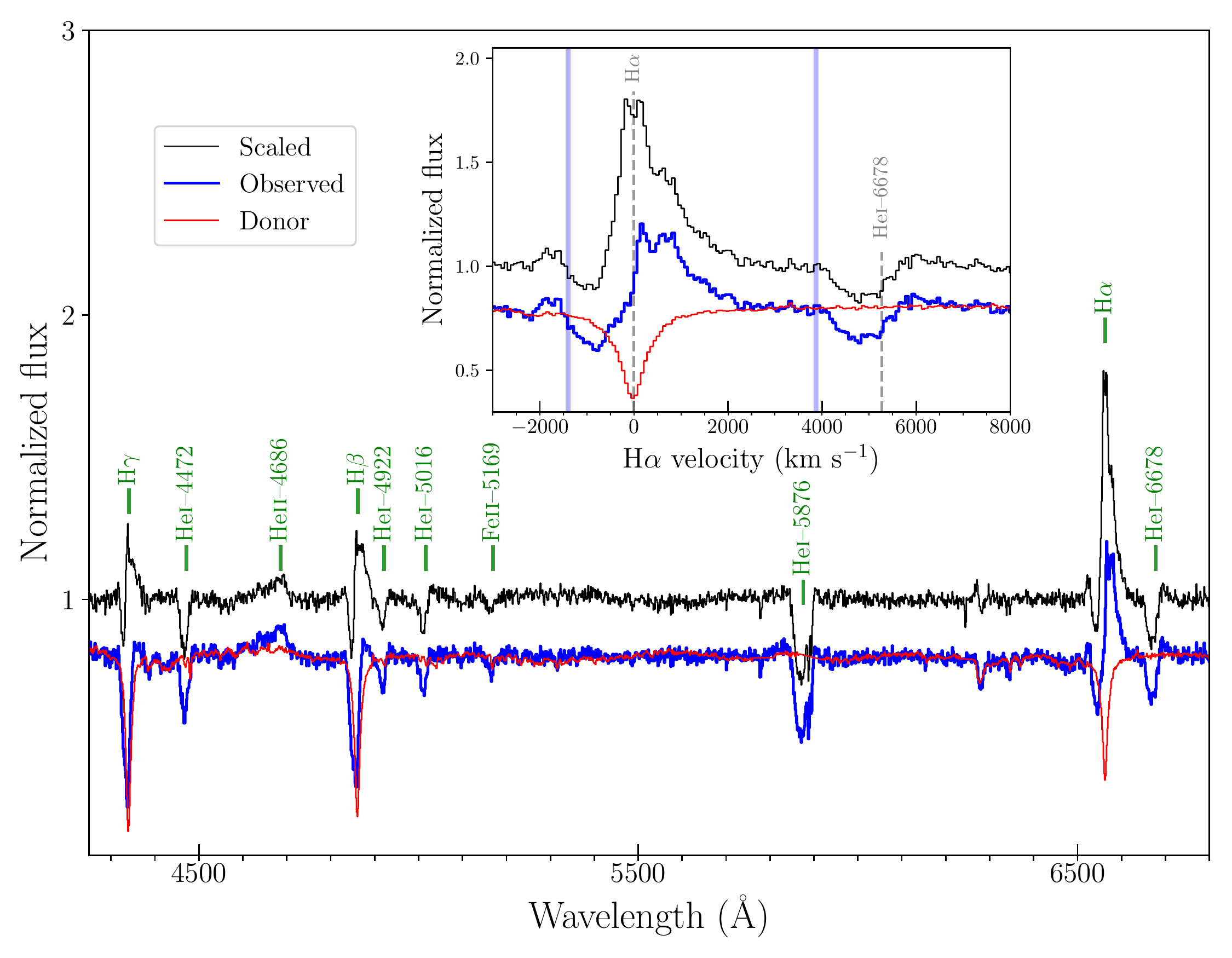}
\caption{2002 spectrum (Jun 12) in the same format as in Fig. \ref{fig:1999}.
The inset shows the \ha\ and \hei--6678 transitions, respectively (dashed, vertical lines). The vertical, blue bands mark the terminal velocity of the wind as measured in the P-Cygni absorption ($1400~\kms$). }
\label{fig:2002}
\end{center}
\end{figure*}

\subsection{Photometry}
In order to properly interpret the spectroscopic data, we included in the analysis the photometric database  presented in \citet{Uemura2002, Uemura2004b, Uemura2005} for the 1999, 2002, and 2004 outburst, respectively, where we refer the reader for further details. It includes data in both the R and the V Johnson bands; the latter collected via the variable star network VSNET. Fig. \ref{fig:lc} shows the optical light-curves for the three outburst included in this work, together with the epochs (green bands) at which the optical spectra were obtained. It is clear from this figure that our spectroscopic data were taken after the main activity periods, sometimes during epochs with small-scale flaring activity.

\section{Analysis and results} 
\label{analysis}

\subsection{1999 outburst}
\label{1999}
The 1999 outburst is the brightest outburst ever observed of V4641 Sgr with a flux peak of $\sim 12$ Crab in the soft X-ray band \citep{Smith1999} and V $\sim$ 10. As shown in Fig. \ref{fig:lc}, our first spectrum (Sep 17) was taken only $\sim 1.5$ days after the optical peak, but following a dramatic drop of $\sim 3$ magnitudes over this lapse of time. This spectrum is characterised by strong and broad \ha\ emission (blue line in Fig.  \ref{fig:1999}), whilst the rest of the Balmer series is observed in absorption, owing to the strong contribution of the donor (red line in Fig.  \ref{fig:1999}).  However, the donor-scaled spectrum (black line in Fig.  \ref{fig:1999}) reveals clear \hb, \hg, and \hd\ emission line profiles suggesting that they are due to the presence of accretion disc emission lines, which are typical of BHTs in outburst (but see section \ref{odata}). The  September 17 \ha\ line profile is complex and deserves particular attention:
\begin{itemize}

\item It has an equivalent width \footnote{the equivalent width is defined as positive for emission lines} of $70 \pm 0.2$ \AA\ (the individual spectra show values in the range of 72--65 \AA) and a full-width-at-zero-intensity of $\sim$ 6000 \kms. This latter value is very similar to that observed in the June 2015 outburst of V404 Cyg during the so-called nebular phase. It was also witnessed following a sharp 3-magnitude drop from the outburst peak, and most likely reflects the optically thin phase of an expanding nebulae created by the strong accretion disc wind detected in this system (\citetalias{Munoz-Darias2016}; see also \citealt{Rahoui2017}). Near-infrared spectroscopy simultaneous with our September 17 spectrum \citep{Charles1999,Chaty2003} also reveals broad \hei\ and Br$\gamma$ emission lines, extending to the same velocity (i.e. $\pm 3000$ \kms) than that observed in \ha\ (Fig.  \ref{fig:1999}).

\item The line profile is very complex, a property also observed in other outbursts of the source \citep[e.g.][]{Lindstrom2005}.  In particular, clear absorption troughs are observed at $-1300$ \kms and $-2200$ \kms, but both features match wiggles present in the quiescence spectrum (red line in Fig.  \ref{fig:1999}). The latter throat is most likely due to a forest of telluric lines present at 6514--21 \AA\ \citep{Kurucz2006}, while   the origin of the former ($\sim$ 6532--38 \AA) is less certain and it might be associated with a diffuse interstellar band centred at 6534~\AA\ \citep{Hobbs2009}. We note that observations were performed at high airmass (2--3) favouring contamination by telluric lines. On the other hand, we can distinguish a hump in the red wing placed at $\sim 1000$ \kms\ with respect to \ha.
\end{itemize} 
The September 18 spectrum shows much weaker \ha\ emission (equivalent width of $2.8 \pm 0.1$ \AA), which is heavily contaminated by the donor. Indeed, no \hb\ nor \hg\ emission is recovered in the donor-scaled spectrum.

\subsection{2002 outburst}
\begin{figure}
\begin{center}
\includegraphics[keepaspectratio,width=8.cm]{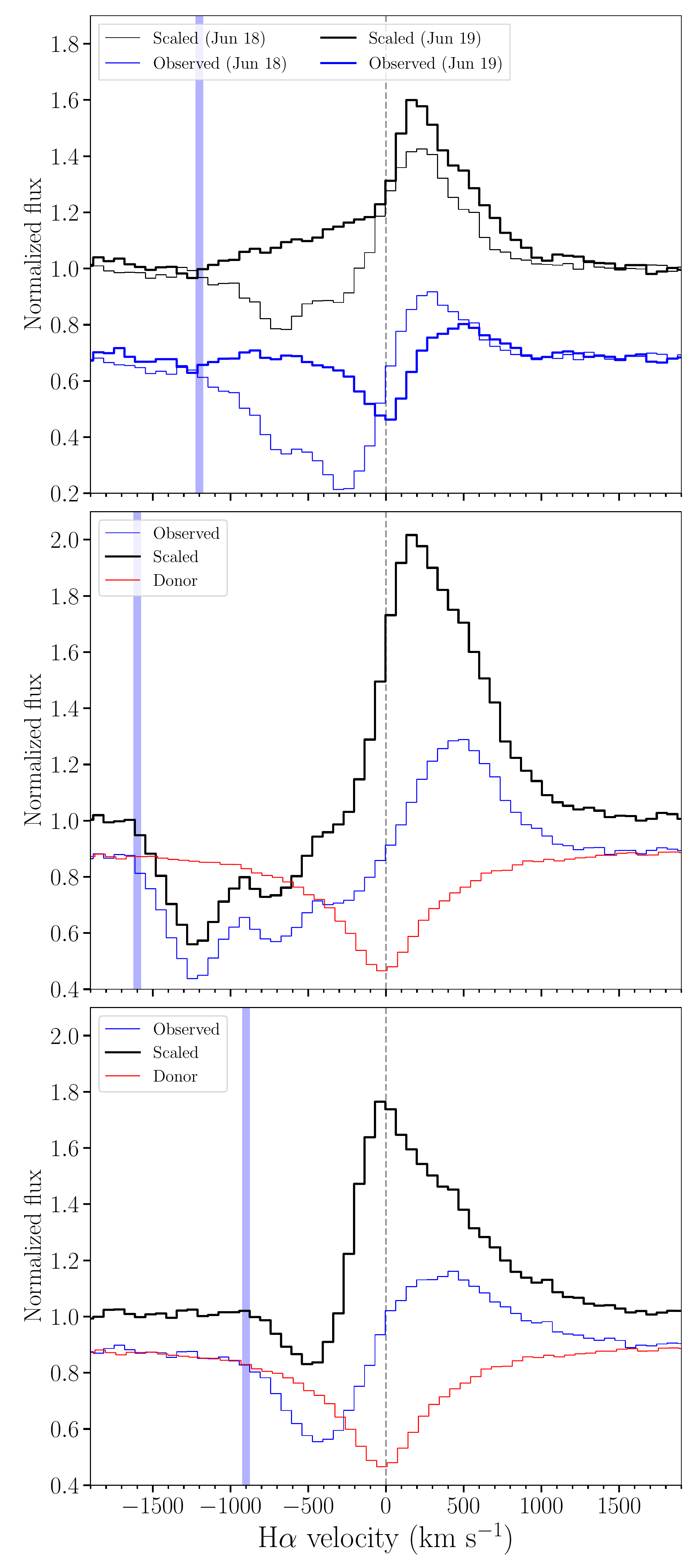}
\caption{Examples of \ha\ outflows using both observed and donor-scaled, normalized spectra. \textbf{Top panel:} 2002 outburst. The Jun 18 data with a P-Cygni absorption is shown as a thin line while a blue wing reaching the same velocity (1200 \kms) is observed the next day (Jun 19).  Observed spectra (blue solid line) are offset by -0.3 in the Y-axis. \textbf{Middle panel:} complex, blue-shifted absorptions (observed on Jul 4, 2002) showing a terminal velocity of 1600 \kms. \textbf{Bottom panel:} 2004 outburst. P-Cyni profile with a terminal velocity of 900 \kms\ observed on Jul 21. For every panel the colour code is the same as in Figs \ref{fig:1999} and \ref{fig:2002}. The vertical blue bands mark the wind terminal velocity for each case.}
\label{fig:mspec}
\end{center}
\end{figure}

We obtained an extensive coverage during the 2002 outburst with 18 epochs of spectroscopy over 40 days (see Table \ref{log} and Fig. \ref{fig:lc}). As it happened with the 1999 event, we missed the outburst peak, but the photometric data available indicates that the system displayed several flaring episodes during  the above time interval.  The \ha\ line shows positive equivalent widths (i.e. emission) only during 3 out of the 18 epochs, but this number rises to 14 in the donor-scaled database. The remaining 4 spectra show weak emission lines and blue-shifted absorptions, which are likely wind-related. The strongest \ha\ emission is detected during our first spectrum (equivalent width of $9.3\pm0.2$ \AA), which was obtained only 0.5 days after a 2-magnitude flare (mid panel in Fig \ref{fig:lc}).  However, the most interesting data were taken from Jun 12 onwards, when pronounced P-Cygni profiles are detected.  For these profiles the wind terminal velocity is taken as the velocity measured from the left edge of the blue absorption component. We estimate this method to be accurate within 100 \kms.
\begin{itemize}
\item On Jun 12 we observed the strongest P-Cygni profiles, with deep, blue-shifted absorption components in the Balmer and \hei\ emission lines (Fig. \ref{fig:2002}). The same feature is also present in Fe~{\sc ii} at 5169 \AA.  We note that both Fe~{\sc ii} and the \hei\ transitions are not contaminated by companion star-related features and therefore the observed and donor-scaled line profiles are identical. All the above lines show a terminal velocity of $\sim 1400$ \kms\ in the blue-shifted absorption component, while virtually no emission is present in  \hei\ and Fe~{\sc ii}. This might indicate that the geometry of the outflow deviates from spherical, resembling the blue-shifted absorptions typically observed in hot X-ray winds, where the geometry is thought to be equatorial \citep{Ponti2012}. Clear emission is present in the Balmer lines when looking at the donor-scaled spectrum, but this has likely a strong accretion disc contribution. Finally, we note that both \heii\ at 4686 \AA\ and the nearby Bowen complex ($\sim 4640$ \AA) are clearly detected with a summed equivalent width of $6.7\pm0.3$ \AA. Their presence is interesting since these higher excitation emission lines were notoriously absent during the phases where similarly strong P-Cygni profiles were detected in V404 Cyg (\citetalias{Munoz-Darias2016}). 

\item Weaker P-Cygni profiles pointing to wind terminal velocities in the range of $900$--$1600$ \kms\ are observed from Jun 14 to July 11.  These profiles are sometimes very complex, with low and high velocity components as shown in the middle panel of Fig. \ref{fig:mspec}. As it happened on Jun 12, the Jun 17 observation shows both blue-shifted absorptions (only in \hei; terminal velocity of $\sim 900$ \kms) and strong \heii--Bowen emission (equivalent width of $7.0\pm0.2$). 

\item On Jun 18 a P-Cygni profile is observed in \ha\ indicating a wind terminal velocity of $\sim 1200$ \kms. The next day (Jun 19) the same emission line shows no absorption but a blue wing extending up to precisely the same velocity (top panel in Fig. \ref{fig:mspec}). This can be interpreted as the transition from optically thick to optically thin ejecta (i.e. to a nebular phase). Tentative, optically thin wind phases (i.e. extended wings) are also observed on Jul 7.

\end{itemize}

\subsection{2004 outburst}
Two single spectra were obtained on July 19 and 21, 2004. These observations were taken about 2 weeks after the peak of the outburst, which was followed by a sharp decay. Fig. \ref{fig:lc} shows $\sim$0.5 mag--flaring episodes  occurring contemporaneously with our observations (we note that data are not simultaneous). On Jul 21, \ha\ is clearly detected in emission and shows a blue-shifted absorption component pointing to a wind terminal velocity of 900 \kms (bottom panel in Fig. \ref{fig:mspec}). The donor-scaled spectrum  reveals P-Cygni shapes at this velocity for the \hb\ and \hg\ lines as well (not shown). Two days before, on Jul 19, we detected weak \ha\ emission that reveals a complex P-Cygni profile with two absorption components at $\sim 1000$ and $\lesssim 500$ \kms. The former component is also present in \hb\ and \hg\ in the donor-scaled spectrum, while for the latter we cannot rule out contamination from the donor star absorption line. Besides the Balmer components, only a weak \hei\ emission line at 6678 \AA\ is detected in the Jul 19 spectrum and it does not show any apparent wind-related signature. A weak $\sim 600$ \kms P-Cygni absorption component was reported by \citet{Lindstrom2005} during the rise of the outburst (15 days before our observations) in Fe~\textsc{ii}. They also detected broad \ha\ emission with wings extending up to $\pm 2500$ \kms\, and therefore comparable with our 1999 result (Fig. \ref{fig:1999}). 

\section{discussion}
\label{discussion}

We have detected conspicuous spectral features (e.g. Figs. \ref{fig:1999} and \ref{fig:2002}) indicating the presence of wind-type outflows in the BHT V4641 Sgr. The winds show terminal velocities up to $\sim 1600$ \kms and were present in the 1999, 2002 and 2004 outbursts of the BHT V4641 Sgr. This is inferred from (P-Cygni) blue-shifted absorptions in H, \hei and Fe~{\sc ii}, but this range extends up to $\sim 3000$ \kms\ when considering the broad \ha\ emission present during our 1999 observation, which was taken 1.5 days after the extremely bright peak of the outburst. The most likely origin for the wind is the accretion disc of the system. We note that even when the companion star is hotter (B9-III with $\sim 10500$ K) than the companions usually found in BHTs \citep{Orosz2001, MacDonald2014}, it cannot be the source of the observed outflows. Late B-type stars can show relatively weak and slow winds, but H and \hei\ emission lines are only seen in B supergiants (i.e. luminosity class I). Indeed, winds with terminal velocities $\gtrsim 1000$ \kms are only present in supergiants with effective temperatures $\gtrsim$ 20000 K and masses above $\gtrsim 20$\sol. Luminosity classes II, III and V only show the observed wind velocities for temperatures above $\sim 30000$ K (i.e. the O-type regime; see \citealt{Kudritzki2000} for a review in stellar winds; see also \citealt{Cox2000}). Heating effects might modify the inner side temperature of the companion, but these should not have a significant effect in this matter.  For instance, \citet{Orosz2014} only found a difference of $\sim 1500$ K between the irradiated and non-irradiated sides of the B-type companion in the persistent black-hole binary LMC X-3.  In addition, we detected wind features at fluxes much lower than the outburst peak where irradiation effects are likely much lower than in the previous case. On the other hand, the wind velocities are very similar to those seen in accretion disc winds in low-mass X-ray binaries and the emission lines that we detect are characteristic of the accretion discs of these objects.   

The present work is, to the best of our knowledge, the first systematic study focussed on the detection and characterization of wind-related features in this system. However, the presence of similar signatures in a few optical and infrared spectra of V4641 Sgr has been reported, and interpreted in the same way, in the past. In particular, broad \ha\ emission with wings extending beyond $\pm~2000~\kms$ was observed in the 1999  and 2004 outbursts \citep{Chaty2003, Lindstrom2005}. In addition, a weak, low-velocity P-Cygni profile was witnessed in 2004 \citep{Lindstrom2005}. 

We have used 39 spectra taken over 22 epochs to show that accretion disc wind-related features are almost ubiquitously present during, at least, the faint phases following the bright outburst peaks of this source. These features can be remarkably conspicuous, such as in the case of our Jun 12, 2002 data (Fig. \ref{fig:2002}) where deep, blue-shifted absorptions are observed in \hei, Fe~{\sc ii}  and the Balmer series (H).  We note that the strong contribution of the donor during the outburst complicates the analysis of the H transitions. This means that even the \ha\ line parameters, such as the equivalent width, must be taken with caution when compared with those from other systems. Wind velocities, on the other hand, are free from these uncertainties, as it is shown in the cases where it can be simultaneously measured in both \hei\ (i.e. not contaminated by the donor) and H lines (e.g. Fig. \ref{fig:2002}). 

\subsection{The accretion-ejection coupling}
The intimate but increasingly complex relation between accretion and ejection processes in accreting black holes and neutron stars has been a matter of strong debate and research for several years \citep[e.g.][for a review]{Fender2016}. In light of the detection of optical accretion disc winds presented here, it is important to consider the X-ray and radio properties  of the source.  However, the bright phases of the outbursts of V4641~Sgr are typically very short and therefore making follow-up at multiple wavelengths is not trivial.  

\subsubsection{X-ray properties}
Besides the obvious $\sim 12$ Crab X-ray peak of the 1999 outburst \citep{Smith1999}, and some activity during the most active phase of 2002, only sporadic and faint detections were recorded by the All-Sky-monitor on-board the \textit{Rossi X-ray Timing Explorer (RXTE)}. On the other hand, the few dedicated high signal-to-noise observations of the source are characterised by strong and broad emission above $\sim$ 4 keV, which is rather peculiar for a BHT. This has been interpreted as resulting from obscuration and reprocessing from dense material surrounding the black hole. This kind of X-ray spectrum was firstly observed during the large 1999 flare, which was probably super-Eddington \citep{Revnivtsev2002}, but it is also present in the  2003 and 2004 outbursts at much lower X-ray luminosities \citep{Maitra2006, Gallo2014,Morningstar2014}. The 2004 data, obtained by \textit{Chandra} on Jul 17 (i.e. only 4 days before our optical spectroscopy, but 10 days after the sharp outburst peak) can be fitted with high values of X-ray absorption, likely approaching the Compton-thick regime,  and large covering factors \citep{Morningstar2014}. This model is also proposed by \citet{Maitra2006}, whilst \citet{Gallo2014} suggest that the complex spectral behaviour results from highly Doppler boosted Fe emission from a jet, resembling that observed in blazars. From the timing point of view, the few observations with high count-rates show a significant amount of fast variability and rapid changes in the hardness of the system. The \textit{RXTE} observation at the peak of the 1999 outburst is characterized by high levels of aperiodic variability \citep{Wijnands2000}, a property that unambiguously characterises the hard accretion states of BHTs \citep[e.g.][]{Munoz-Darias2011}.  Similar observations taken over the 2003 outburst during much fainter epochs also show this behaviour \citep{Maitra2006}. From the spectral point of view, all the above works suggest, within the complexity of the spectral fitting in this source, photon indices lower than $\sim 2$. All in all, and considering the low luminosity phases at which our spectra were taken, it seems very likely that the system was in the hard state during most, if not all, the observations where we detect the outflowing wind.  

\subsubsection{Radio emission}
V4641 Sgr is a very interesting source with respect to the radio emission. In particular, and as happened with the X-rays and the optical emission, a strong ($\sim 0.4$ Jy) radio flare was detected during the 1999 outburst (Sep 16; \citealt{Hjellming2000}). The jet was resolved, with an approaching component showing apparent superluminal velocities. The decay of this event was observed for $\sim 2$ more days, while a stationary component was detected for a month. The September 17, 1999 detection of extremely broad \ha\ (Fig. \ref{fig:1999}), \hei\ and Br$\gamma$ (\citealt{Chaty2003}) emission is therefore simultaneous with radio jet emission. However, the exact time of the  ejection is unknown, and, in principle, the jet could be described by the standard optically-thin decay of a single event \citep{vanderLaan1966}. Likewise, our favoured interpretation for the broad emission lines is a so-called nebular phase of a previously emitted wind-like ejecta. Therefore, for the 1999 event the jet-wind simultaneity seems likely but there is not definitive proof. In the 2002 and 2004 outbursts, we detected conspicuous wind features during several epochs at relatively low luminosity, most likely during the hard state. In agreement with this, compact radio emission was detected by the \textit{Very Large Array} at the times of our wind detections \citep{Rupen2002,Rupen2004}. In 2002, the jet is detected (with an inverted spectrum characteristic of the hard state) on Jun 13 and Jun 24, and we have confirmed wind detections on e.g. Jun 12 and Jun 14 (see Table \ref{log}). Interestingly, no radio emission is detected on Jun 5, and optical wind is not observed either. In 2004, the radio light-curve (see links in \citealt{Rupen2004}) shows 5 GHz emission at the level of 1--3 mJy during Jul 18, Jul 19 and Jul 20, while the source was not detected (<0.4 mJy) on Jul 23. Optical wind was detected on Jul 19 and Jul 21. We conclude that, at least for the 2002 and 2004 outbursts, there is strong evidence for the presence of an optical wind simultaneous with the radio jet during the hard state.

\subsection{The inevitable comparison with V404 Cyg}

Despite the very significant difference in companion star type, with V404 Cyg harbouring a K sub-giant, rather typical in BHTs \citep{Casares1992, King1993,Munoz-Darias2008}, and V4641 Sgr a late B-type donor \citep{Orosz2001}, the phenomenology observed in both objects have several ``unique" aspects in common.  In the first place, they are able to power luminous outbursts that likely exceed the Eddington limit during some phases \citep[e.g.][]{Revnivtsev2002, Motta2017a}. However, the really active phases of theses events can be very short and they do not show steady disc dominated (i.e. soft) states, but hard X-ray spectra from which high levels of photometric absorption are derived. Indeed, since there is strong evidence for significant amount of expelled mass surrounding the system, scenarios similar to those thought to occur in obscured (i.e. Type-2) active galactic nuclei have been proposed  \citep[e.g.][]{Zycki1999, Morningstar2014, Motta2017b}. On the other hand, both sources are radio loud during their accretion episodes, strongly suggesting that they do not go beyond intermediates states during their bright phases. 

\citetalias{Munoz-Darias2016} proposed an scenario in which the strong wind detected in the V404 Cyg might regulate the outburst evolution. The idea is that only the innermost part of the accretion flow, which would not be affected by the wind, is accreted, explaining the short duration of the bright outburst phases. In principle, the same scenario would be able to qualitatively account for the behaviour of V4641 Sgr (see below), albeit we would need to infer the amount of mass ejected. To obtain this number is beyond the scope of this paper since it would require proper modelling and effectively deal with the companion star contribution to the optical spectra. However, we note that it might be very high if we consider the conspicuous wind features and the large amounts of absorbing material detected in X-rays. It is important to bear in mind that there is evidence suggesting that disc winds might carry a very significant amount of the mass being accreted (e.g. \citealt{Neilsen2011, Neilsen2013, Ponti2012, Fender2016}, \citetalias{Munoz-Darias2016}, \citealt{Tetarenko2018}). We note that this wind instability scenario has been proposed to explain the behaviour of the ultraluminous X-ray source HLX-1 \citep{Soria2017}.   

\subsubsection{On the optical wind mechanism and its relation with other outflow phenomena}
While the vast majority of the hot X-ray wind detections were found during soft states (note that the state classification is only qualitative in several cases), that is, when the jet is not observed \citep[e.g.][]{Neilsen2009,Ponti2012}, the V404 Cyg observations clearly showed that accretion disc winds and jets can coexist (\citetalias{Munoz-Darias2016},~ \citealt{Munoz-Darias2017}) and further evidence was presented in \citet{Homan2016}. The results presented in this paper provide additional evidence for this picture and show that hard state, cold (i.e. optical/infrared) winds simultaneous with the radio jet do occur, and at least for these two systems seem to be very common. This raises several important questions, such as the connection between optical and X-ray winds and whether or not they are different faces of the same phenomena or just different outflows from e.g. different regions of the accretion flow. In this regard, \citet{Bianchi2017} showed that the apparent absence of hot X-ray winds in the hard state (once they have been observed in the soft state) could be related to photoionization instability (see also \citealt{Chakravorty2013}). Interestingly, one of the hard state allowed solutions is a cold ($\sim$ neutral) phase of the wind (the other is a very hot phase). The nature and characteristics of this tentative cold outflow needs further investigation.

The properties of the optical wind observed in V4641 Sgr are very similar to those of V404 Cyg. Both are observed predominately in H and \hei\ recombination lines, show velocities in the range $\sim 1000$--$3000$ \kms and display at least two distinctive phases, as implied by the detection of P-Cygni/blue-shifted absorptions and strong emission lines with broad wings. There is a likely causal connection between these two phases, with the latter, so-called nebular phase, being the expanding (optically-thin) phase of a previously launched and thicker ejecta that produces the former phase (we direct the reader to \citealt{MataSanchez2018} where a study on the relation between these two wind phases is presented). MD16 showed that the wind features observed in V404 Cyg are compatible with the Compton-heated wind mechanism (thermal wind; \citealt{Begelman1983}), and thus, given the very similar observables, this scenario could be also applied to V4641 Sgr. Following MD16, and using a wind terminal velocity of $1500~\kms$ and  10 M$_{\odot}$ as the black hole mass, we obtain a wind launching radius of 2 light-seconds, while the disc has a size of $\sim 30$ and $\sim 17$ light-seconds for V404 Cyg and V4641 Sgr, respectively. MD16 also showed that at this launching radius the disc temperature (from \citealt{Shakura1973}) is consistent with the H and He features present in the optical spectra for luminosities $\lesssim 0.1$ times the Eddington luminosity.
 Therefore, these thermal winds would be launched from the ``outer" disc, but they could still isolate an inner, relatively small, region from the rest of the accretion flow, preventing the black hole from being fed for much longer time (i.e. longer outbursts). Nevertheless, there is no reason to think that alternative wind-mechanism scenarios could equally explain part or all the observed phenomenology. 

The simplest scenario might be a radiation driven wind produced when the systems approach the Eddington limit. Despite that there is always uncertainty on the actual luminosity of the source because of the strong, variable absorption, optical wind is observed at stages with low accretion activity at low (X-ray, optical and radio) luminosity. It is important to bear in mind that wind was detected even in the sequel 2015 outburst of V404 Cyg, which was much less extreme than e.g. the 2015 main outburst or the 1999 outburst of V4641 Sgr. Thus, radiation driven cannot be the main/only wind process at work. However, it might explain the nebular phase of the 1999 outburst, given that it was observed only 1.5 days after the very luminous outburst peak (\citealt{Revnivtsev2002}).  
Another common outflow mechanism is line-driven wind, which can be effective at luminosities as low as $\sim 10^{-3}$ the Eddington luminosity. However, it is usually not considered in X-ray binaries since X-ray emission from the disc would over-ionize the wind, unless the UV-emitting region is shielded \citep{Proga2002}. In this regard, a scenario in which the inner disc surrounds the central source and produces high X-ray absorption (as e.g. proposed by \citealt{Revnivtsev2002}  for V4641 Sgr and \citealt{Motta2017b} for V404 Cyg) might yield the required shielding and suitable (i.e. less energetic) photons. Again, it is unclear whether this accretion geometry is valid for the low X-ray, optical and radio emission phases with optical wind detections in both V404 Cyg and V4641 Sgr.  Finally, we note that winds launched by the accretion disc magnetic field (i.e. magnetic winds) have been proposed to explain X-ray observations of BHTs (e.g. \citealt{Miller2006}). This scenario is even more difficult to test, but it might be valid for both V404 Cyg and V4641 Sgr.

\section{Conclusions}
We have presented compelling evidence for the presence of optical winds in the 1999, 2002 and 2004 outbursts of the black hole X-ray binary transient V4641~Sgr, which become the second source of this class, together with V404 Cyg, where the presence of these cold outflows has been clearly established. The wind is detected in H (Balmer) and \hei\ lines in the form of both P-Cygni/blue-shifted absorption components and strong emission lines with broad wings. The terminal velocity of the wind, as inferred from the blue-shifted absorptions, is typically slightly above $\sim 1000$ \kms, although lower velocities are observed (and reported in the literature; \citealt{Lindstrom2005}). On the other hand, the nebular phase observed after the likely super-Eddington 1999 flux peak indicates outflow velocities of up to  $\sim 3000$ \kms.  As for the case of V404 Cyg, most of the wind observations are obtained during low activity epochs characterised by faint X-ray, optical and radio emission; therefore simultaneous with the radio-jet and during the hard state.  Future observations of this object during forthcoming outbursts (e.g. detailed X-ray spectroscopy), as well as new optical/infrared spectroscopic studies of active black hole transients should shed more light on the nature of these cold outflows, their impact on the accretion process and their relation with other accretion/ejection phenomena.



\section*{Acknowledgements}

TMD and MAPT acknowledge support by the Spanish MINECO under grant AYA2017-83216-P. TMD and MAPT acknowledge support via Ram\'on y Cajal Fellowships RYC-2015-18148 and RYC-2015-17854, respectively. We are thankful to Makoto Uemura for providing the photometric data presented in \citet{Uemura2002, Uemura2004b, Uemura2005}. TMD is thankful to Sergio Sim\'on-D\'iaz for useful discussion on the properties of B stars. {\sc molly} software developed by Tom Marsh is gratefully acknowledged.




\bibliographystyle{mnras}
\bibliography{/Users/tmd/Dropbox/Libreria} 



%


\bsp	
\label{lastpage}
\end{document}